\documentclass[english,aps,preprint]{revtex4}
\usepackage{pslatex}
\usepackage[T1]{fontenc}
\usepackage[latin1]{inputenc}
\usepackage{graphicx}

\makeatletter

\usepackage{babel}
\makeatother
\begin{document}

\preprint{Dec 7, 2005}

\title{Capacitance Calculations Using the Lattice Green Function in Two
Dimensions}

\author{Stefan Hollos}

\email{stefan@exstrom.com}

\homepage{http://www.exstrom.com/stefan/stefan.html}

\affiliation{Exstrom Laboratories LLC, 662 Nelson Park Dr, Longmont, Colorado
80503, USA.}

\author{Richard Hollos}

\affiliation{Exstrom Laboratories LLC, 662 Nelson Park Dr, Longmont, Colorado
80503, USA.}

\begin{abstract}
We show how to use the lattice Green function to calculate capacitances
in two dimensions with boundary conditions at infinity. It is shown
how to calculate coefficients of capacitance and induction from the
lattice Green function. A general analysis of two arbitrary conductors
is carried out. It is shown how the calculations can be simplified
in the case of identical conductors when certain symmetry conditions
are met. Example calculations for a parallel and coplanar stripline
are shown. The use of the two conductor formulas for the case of three
or more conductors is discussed.
\end{abstract}
\maketitle

\section{introduction}

In a two dimensional homogeneous and unbounded space, the Poisson
equation and its solution are given by\begin{equation}
\nabla^{2}\phi(\vec{r})=-\frac{\rho(\vec{r})}{\epsilon}\label{eq:1}\end{equation}
\begin{equation}
\phi(\vec{r})=\frac{1}{2\pi\epsilon}\int\ln\left(\frac{r_{0}}{\vert\vec{r}-\vec{r}'\vert}\right)\rho(\vec{r}')\; d\vec{r}'\label{eq:2}\end{equation}
When the charge density is known then eq. \ref{eq:2} reduces the
problem of finding the potential at any point in space to a simple
integration. A more interesting and difficult problem occurs when
the charge density is not known. An example of such a probelm is the
case of two conductors each of which is held at some constant potential.
Eq. \ref{eq:2} is then an integral equation for the charge density
on the conductors and it is difficult to solve analytically in all
but a few simple cases. Many approximation schemes have been devised
to deal with this problem and in most cases the solution must be found
numerically. One commonly used approach is to discretize eq. \ref{eq:2},
thereby turning it into a matrix equation, which can then be solved
using standard techniques. The disadvantage of this approach is that
it is an approximation of the continuous as well as the discrete version
of the problem, i.e. it is an approximation in both continuous and
discrete space.

An alternative approach is to formulate the entire problem in a discrete
space. Physically the model is an infinite square lattice with capacitors
connecting the nodes. Sets of adjacent nodes, that correspond to discretized
versions of conductors, are held at constant potential and the problem
is to find the charges at those nodes. Mathematically this approach
is equivalent to replacing the Laplacian in eq. \ref{eq:1} with its
finite difference approximation. The discrete space version of eq.
\ref{eq:1} is then\begin{equation}
\sum_{m}L_{nm}\phi(\vec{r}_{m})=-\frac{\lambda(\vec{r}_{n})}{\epsilon}\label{eq:3}\end{equation}
where $\vec{r}_{n}$ is a lattice vector of the form: $\vec{r}_{n}=n_{1}\vec{a}_{1}+n_{2}\vec{a}_{2}$,
$n_{i}=$integer, $\vec{a_{i}}=a\hat{x}_{i}$ and $a=$lattice spacing.
$\lambda(\vec{r}_{n})$ is the linear charge density at node $\vec{r}_{n}$.
$L_{nm}$ is a matrix element of the lattice Laplacian and is defined
as\begin{equation}
L_{nm}=-4\delta(\vec{r}_{n},\vec{r}_{m})+\sum_{i=1}^{2}\left[\delta(\vec{r}_{n}+\vec{a}_{i},\vec{r}_{m})+\delta(\vec{r}_{n}-\vec{a}_{i},\vec{r}_{m})\right]\label{eq:4}\end{equation}
To simplify the notation we will write eq. \ref{eq:3} in the following
form\begin{equation}
\sum_{m}L_{nm}\phi(\vec{r}_{m})=-q(\vec{r}_{n})\label{eq:5}\end{equation}
where $L_{nm}$ now has units of capacitance and $q(\vec{r}_{n})$
is a charge at node $\vec{r}_{n}$. The solution of this equation
for $\phi(\vec{r}_{n})$ is\begin{equation}
\phi(\vec{r}_{n})=\sum_{m}G_{nm}\, q(\vec{r}_{m})\label{eq:6}\end{equation}
where $G_{nm}$ is a matrix element of the Green function with units
of elastance. The Green function in this problem is defined as, $G=-L^{-1}$,
where $L$ is the lattice Laplacian operator. Eq. \ref{eq:6} is the
discrete space version of eq. \ref{eq:2}.

$G_{nm}$ is a function of $\vert\vec{r}_{n}-\vec{r}_{m}\vert$ only,
so a more convenient notation for eq. \ref{eq:6} is\begin{equation}
\phi(n_{1},n_{2})=\sum_{m_{1},m_{2}}G(p_{1},p_{2})\, q(m_{1},m_{2})\label{eq:7}\end{equation}
where $p_{1}=\vert n_{1}-m_{1}\vert$ and $p_{2}=\vert n_{2}-m_{2}\vert$.
The problem with this equation however, is that $G(p_{1},p_{2})$
is infinite for all values of $p_{1}$ and $p_{2}$. This same problem
occurs in the continuous case, eq. \ref{eq:2}, when the reference
point for the potential, $r_{0}$, is taken to infinity. The way around
this problem is to use $g(p_{1},p_{2})=G(0,0)-G(p_{1},p_{2})$ which
is finite for all values of $p_{1}$ and $p_{2}$ \cite{hollos05_1,hollos05_3}.
Eq. \ref{eq:7} then becomes\begin{equation}
\phi(n_{1},n_{2})=-\sum_{m_{1},m_{2}}g(p_{1},p_{2})\, q(m_{1},m_{2})\label{eq:8}\end{equation}
This equation will solve the original problem as long as all the charges
sum to zero. We will now show how to use this formalism to calculate
the capacitance between conductors in two dimensions.

\section{general analysis of two conductors}

Consider the case of two conductors discretized so that there are
$n_{1}$ charges on conductor 1 and $n_{2}$ charges on conductor
2. The conductors are held at constant potentials $\phi_{1}$ and
$\phi_{2}$. The equation for this system is written as follows\begin{equation}
-\left(\begin{array}{cc}
\mathbf{G}_{11} & \mathbf{G}_{12}\\
\mathbf{G}_{21} & \mathbf{G}_{22}\end{array}\right)\left(\begin{array}{c}
\vec{\mathbf{q}}_{1}\\
\vec{\mathbf{q}}_{2}\end{array}\right)=\left(\begin{array}{c}
\phi_{1}\vec{\mathbf{e}}_{1}\\
\phi_{2}\vec{\mathbf{e}}_{2}\end{array}\right)\label{eq:9}\end{equation}
$\vec{\mathbf{q}}_{i}$ is an $n_{i}$ dimensional vector of the charges
on conductor $i$. $\vec{\mathbf{e}}_{i}$ is an $n_{i}$ dimensional
vector with all elements equal to 1. $\mathbf{G}_{ii}$ is an $n_{i}\, x\, n_{i}$
matrix that gives the contribution of the charges on conductor $i$
to the potential of the conductor. $\mathbf{G}_{ij}$ is an $n_{i}\, x\, n_{j}$
matrix that gives the contribution of the charges on conductor $j$
to the potential of conductor $i$. The elements of $\mathbf{G}_{ij}$
depend only on the absolute separation of two charges therefore it
will always be true that $\mathbf{G}_{ij}=\mathbf{G}_{ji}^{T}$.

As a simple example, take two conductors discretized such that there
are two charges on both conductors. Let the charges on conductor 1
be at $(0,0)$ and $(0,1)$ and the charges on conductor 2 be at $(2,0)$
and $(3,0)$. The $\mathbf{G}_{ij}$ submatrices in eq. \ref{eq:9}
will then be\begin{equation}
\mathbf{G}_{11}=\mathbf{G}_{22}=\left(\begin{array}{cc}
0 & g(1,0)\\
g(1,0) & 0\end{array}\right)\label{eq:10}\end{equation}
\begin{equation}
\mathbf{G}_{12}=\mathbf{G}_{21}^{T}=\left(\begin{array}{cc}
g(2,0) & g(3,0)\\
g(2,1) & g(3,1)\end{array}\right)\label{eq:11}\end{equation}

In general, to calculate the capacitance between two conductors the
matrix in eq. \ref{eq:9} needs to be inverted in order to find the
charges $\vec{\mathbf{q}}_{1}$, and $\vec{\mathbf{q}}_{2}$. The
matrix is easily inverted if it is first factored into a product of
an upper and lower triangular matrix. This can be done in two ways\begin{equation}
\left(\begin{array}{cc}
\mathbf{G}_{11} & \mathbf{G}_{12}\\
\mathbf{G}_{21} & \mathbf{G}_{22}\end{array}\right)=\left(\begin{array}{cc}
\mathbf{I} & \mathbf{0}\\
\mathbf{G}_{21}\mathbf{G}_{11}^{-1} & \mathbf{I}\end{array}\right)\left(\begin{array}{cc}
\mathbf{G}_{11} & \mathbf{G}_{12}\\
\mathbf{0} & \mathbf{G}_{22}-\mathbf{G}_{21}\mathbf{G}_{11}^{-1}\mathbf{G}_{12}\end{array}\right)\label{eq:12}\end{equation}
or\begin{equation}
\left(\begin{array}{cc}
\mathbf{G}_{11} & \mathbf{G}_{12}\\
\mathbf{G}_{21} & \mathbf{G}_{22}\end{array}\right)=\left(\begin{array}{cc}
\mathbf{I} & \mathbf{G}_{12}\mathbf{G}_{22}^{-1}\\
\mathbf{0} & \mathbf{I}\end{array}\right)\left(\begin{array}{cc}
\mathbf{G}_{11}-\mathbf{G}_{12}\mathbf{G}_{22}^{-1}\mathbf{G}_{21} & \mathbf{0}\\
\mathbf{G}_{21} & \mathbf{G}_{22}\end{array}\right)\label{eq:13}\end{equation}
Inverting these upper and lower triangular matrices then leads to
two sets of equations for the submatrices of the inverse. From eq.
\ref{eq:12} we get\begin{eqnarray}
(\mathbf{G}^{-1})_{11} & = & \mathbf{G}_{11}^{-1}+\mathbf{G}_{11}^{-1}\mathbf{G}_{12}(\mathbf{G}_{22}-\mathbf{G}_{21}\mathbf{G}_{11}^{-1}\mathbf{G}_{12})^{-1}\mathbf{G}_{21}\mathbf{G}_{11}^{-1}\label{eq:14}\\
(\mathbf{G}^{-1})_{12} & = & -\mathbf{G}_{11}^{-1}\mathbf{G}_{12}(\mathbf{G}_{22}-\mathbf{G}_{21}\mathbf{G}_{11}^{-1}\mathbf{G}_{12})^{-1}\nonumber \\
(\mathbf{G}^{-1})_{21} & = & -(\mathbf{G}_{22}-\mathbf{G}_{21}\mathbf{G}_{11}^{-1}\mathbf{G}_{12})^{-1}\mathbf{G}_{21}\mathbf{G}_{11}^{-1}\nonumber \\
(\mathbf{G}^{-1})_{22} & = & (\mathbf{G}_{22}-\mathbf{G}_{21}\mathbf{G}_{11}^{-1}\mathbf{G}_{12})^{-1}\nonumber \end{eqnarray}
and from eq. \ref{eq:13} we get\begin{eqnarray}
(\mathbf{G}^{-1})_{11} & = & (\mathbf{G}_{11}-\mathbf{G}_{12}\mathbf{G}_{22}^{-1}\mathbf{G}_{21})^{-1}\label{eq:15}\\
(\mathbf{G}^{-1})_{12} & = & -(\mathbf{G}_{11}-\mathbf{G}_{12}\mathbf{G}_{22}^{-1}\mathbf{G}_{21})^{-1}\mathbf{G}_{12}\mathbf{G}_{22}^{-1}\nonumber \\
(\mathbf{G}^{-1})_{21} & = & -\mathbf{G}_{22}^{-1}\mathbf{G}_{21}(\mathbf{G}_{11}-\mathbf{G}_{12}\mathbf{G}_{22}^{-1}\mathbf{G}_{21})^{-1}\nonumber \\
(\mathbf{G}^{-1})_{22} & = & \mathbf{G}_{22}^{-1}+\mathbf{G}_{22}^{-1}\mathbf{G}_{21}(\mathbf{G}_{11}-\mathbf{G}_{12}\mathbf{G}_{22}^{-1}\mathbf{G}_{21})^{-1}\mathbf{G}_{12}\mathbf{G}_{22}^{-1}\nonumber \end{eqnarray}
In terms of these submatrices the charges on the two conductors are
given by\begin{eqnarray}
\vec{\mathbf{q}}_{1} & = & -\phi_{1}(\mathbf{G}^{-1})_{11}\vec{\mathbf{e}}_{1}-\phi_{2}(\mathbf{G}^{-1})_{12}\vec{\mathbf{e}}_{2}\label{eq:16}\\
\vec{\mathbf{q}}_{2} & = & -\phi_{1}(\mathbf{G}^{-1})_{21}\vec{\mathbf{e}}_{1}-\phi_{2}(\mathbf{G}^{-1})_{22}\vec{\mathbf{e}}_{2}\nonumber \end{eqnarray}
The total charge on conductor $i$ is $Q_{i}=\vec{\mathbf{e}}_{i}^{T}\vec{\mathbf{q}}_{i}$,
so if the first equation is multiplied by $\vec{\mathbf{e}}_{1}^{T}$
and the second equation by $\vec{\mathbf{e}}_{2}^{T}$ then we get
a set of equations relating the total charges on each conductor to
their potentials.\begin{eqnarray}
Q_{1} & = & c_{11}\phi_{1}+c_{12}\phi_{2}\label{eq:17}\\
Q_{2} & = & c_{12}\phi_{1}+c_{22}\phi_{2}\nonumber \end{eqnarray}
The coefficients $c_{ii}$ are known as coefficients of capacitance
and $c_{12}$ is known as a coefficient of induction. The coefficient
$c_{ij}$ is equal to the negative of the sum of all the elements
in the matrix $(\mathbf{G}^{-1})_{ij}$.\begin{equation}
c_{ij}=-\vec{e}_{i}^{T}(G^{-1})_{ij}\vec{e}_{j}\label{eq:18}\end{equation}
From eq. \ref{eq:14} and \ref{eq:15} it is clear that $(\mathbf{G}^{-1})_{ij}=(\mathbf{G}^{-1})_{ji}^{T}$
and therefore $c_{ij}=c_{ji}$. Since the potential and the charge
of a conductor will have the same sign, we have the condition $c_{ii}>0$.
Also since the charge induced by a conductor will have a sign opposite
to the potential of the conductor, we have $c_{12}<0$. 

In addition to eq. \ref{eq:17} we have the requirement that the sum
of all the charges must equal zero, $Q_{1}+Q_{2}=0$. This means that
we can set $Q_{1}=Q$ and $Q_{2}=-Q$ in eq. \ref{eq:17} and solve
for the potentials.\begin{eqnarray}
\phi_{1} & = & \left(\frac{c_{22}+c_{12}}{c_{11}c_{22}-c_{12}^{2}}\right)\, Q\label{eq:19}\\
\phi_{2} & = & -\left(\frac{c_{11}+c_{12}}{c_{11}c_{22}-c_{12}^{2}}\right)\, Q\nonumber \end{eqnarray}
The ratio of the two potentials is then given by\begin{equation}
\frac{\phi_{1}}{\phi_{2}}=-\left(\frac{c_{22}+c_{12}}{c_{11}+c_{12}}\right)\label{eq:20}\end{equation}
The capacitance between the two conductors can then be expressed in
terms of the $c_{ij}$ coefficients as follows.\begin{equation}
C=\frac{Q}{\phi_{1}-\phi_{2}}=\frac{c_{11}c_{22}-c_{12}^{2}}{c_{11}+c_{22}+2c_{12}}\label{eq:21}\end{equation}
Note that the denominator, $c_{11}+c_{22}+2c_{12}$ is equal to the
negative of the sum of all the elements of the inverse of the Green
function matrix in eq. \ref{eq:9} and that the $c_{ij}$'s are functions
only of the geometry of the two conductors. Eq. \ref{eq:21} can be
used to calculate the capacitance between two conductors of arbitrary
size, shape and orientation once the $c_{ij}$ coefficients have been
calculated. Each conductor may consist of more than one disconnected
piece as long as each piece is held at the same potential. We now
look at the special case of two identical conductors, i.e. both conductors
have the same size and shape.

\section{two identical conductors}

For two conductors of the same size and shape the calculation of capacitance
and charge distribution can be simplified in those cases where the
coefficients of capacitance are equal. We begin then by examining
what conditions are required to get $c_{11}=c_{22}$.

Since $c_{ii}$ is equal to the sum of the elements of the $(\mathbf{G}^{-1})_{ii}$
matrix, the two $c_{ii}$ will automatically be equal if the two $(\mathbf{G}^{-1})_{ii}$
matrices are equal. For identical conductors it will always be true
that $\mathbf{G}_{11}=\mathbf{G}_{22}$ so that the expressions for
$(\mathbf{G}^{-1})_{ii}$ in equations \ref{eq:14} and \ref{eq:15}
become\begin{eqnarray}
(\mathbf{G}^{-1})_{11} & = & (\mathbf{G}_{11}-\mathbf{G}_{12}\mathbf{G}_{11}^{-1}\mathbf{G}_{21})^{-1}\label{eq:22}\\
(\mathbf{G}^{-1})_{22} & = & (\mathbf{G}_{11}-\mathbf{G}_{21}\mathbf{G}_{11}^{-1}\mathbf{G}_{12})^{-1}\nonumber \end{eqnarray}
 These expressions are only equal if $\mathbf{G}_{12}=\mathbf{G}_{21}=\mathbf{G}_{12}^{T}$
i.e. $\mathbf{G}_{12}$ must be symmetric.

$\mathbf{G}_{12}$ is a function only of the distance between charges
on the two conductors therefore it can be made symmetric if it is
possible to take one conductor to the other by a reflection about
a horizontal plane, a vertical plane, both a horizontal and vertical
plance, or a diagonal plane. Figure 1 shows the case of two conductors
that can be made congruent through reflection about the vertical plane
$ab$ followed by reflection about the horizontal plane $cd$. The
point $1$ is taken to the point $1'$ so that these charges will
be equal and all charges to the right of point $1$ will equal the
corresponding charges to the left of point $1'$.

With $\mathbf{G}_{12}$ symmetric we have $c_{11}=c_{22}$ and the
equations for the general case simplify. From eq. \ref{eq:20} we
get $\phi_{1}=-\phi_{2}=\phi$ and the potential is related to the
charge as follows\begin{equation}
\phi=\frac{Q}{c_{11}-c_{12}}\label{eq:23}\end{equation}
The capacitance is then given by\begin{equation}
C=\frac{c_{11}-c_{12}}{2}\label{eq:24}\end{equation}
For this symmetric case, each charge on one conductor will have a
corresponding equal and opposite charge on the other conductor so
that in eq. \ref{eq:9} we get $\vec{\mathbf{q}}_{1}=-\vec{\mathbf{q}}_{2}=\vec{\mathbf{q}}$
and the equation reduces to\begin{equation}
(\mathbf{G}_{12}-\mathbf{G}_{11})\vec{\mathbf{q}}=\phi\vec{\mathbf{e}}\label{eq:25}\end{equation}
The total charge on one conductor is then related to the potential
as follows\begin{equation}
Q=\phi\vec{\mathbf{e}}^{T}(\mathbf{G}_{12}-\mathbf{G}_{11})^{-1}\vec{\mathbf{e}}\label{eq:26}\end{equation}
Comparing this equation with eq. \ref{eq:23} gives\begin{equation}
c_{11}-c_{12}=\vec{\mathbf{e}}^{T}(\mathbf{G}_{12}-\mathbf{G}_{11})^{-1}\vec{\mathbf{e}}\label{eq:27}\end{equation}
The capacitance can therefore be calculated from the sum of all the
elements of the inverse of the matrix $\mathbf{G}_{12}-\mathbf{G}_{11}$.

As examples, we will now consider the case of the parallel stripline
and the coplanar stripline. A parallel stripline and the potential
surrounding it is shown in Fig. 2. The upper and lower plates are
at potentials +1 and -1 respectively. A plot of the capacitance as
a function of the ratio of the width of the plates to their separation
is shown in fig. 3. For each ratio the resolution was increased (lattice
constant decreased) until the capacitance value appeared to converge.
Fig. 4 shows the convergence for the ratios 1, 5 and 10 as a function
of the number of charges in the plates. With $r$ equal to the ratio
of the width to the separation, the following equation fits the plot
of the capacitance in fig. 3 with a correlation coefficient of 0.999998.\begin{equation}
C=r+1.12863+0.202069\;\ln(r-0.0964012)\label{eq:28}\end{equation}
This equation agrees well with previous work by Wheeler \cite{wheelerh65}.
The coplanar stripline and the potential surrounding it is shown in
fig. 5. The left and right plates are at potentials -1 and +1 respectively,
with a plate separation of 1/5 the width of a plate. A plot of the
capacitance as a function of the ratio of width to separation of the
plates is shown in fig. 6.

\section{conclusion}

The formalism discussed above can easily be extended to the case of
three or more conductors. The Green function matrix inverse formulas
in eq. \ref{eq:14} and \ref{eq:15} can be applied iteratively in
this case. Start with any two conductors, generate $G_{11}$, $G_{22}$,
$G_{12}$, and then use one of the sets of equations to find the inverse.
This inverse then becomes $G_{11}^{-1}$ in eq. \ref{eq:14} when
the next conductor is included, for which we have a new $G_{22}$
and $G_{12}$, with $G_{12}$ connecting the new conductor with the
previous two. This makes it possible to look at the effect of changes
in the placement of a new conductor (or a single charge) with respect
to a set of conductors without major recalculation efforts.

Another theorem that may be useful in the case of three or more conductors
is Green's Reciprocation Theorem. In its simplest form the theorem
says that if we have a set of $N$ conductors at potentials $\phi_{i}$
and charges $Q_{i}$ ($i=1,\,\ldots\, N$) and then take those same
conductors at new potentials $\phi_{i}'$ and charges $Q_{i}'$ then
the following equation holds\begin{equation}
\sum_{i=1}^{N}Q_{i}\,\phi_{i}'=\sum_{i=1}^{N}Q_{i}'\,\phi_{i}\label{eq:29}\end{equation}
This theorem is easily proven by writing out equations similar to
eq. \ref{eq:17} for the $Q_{i}$ and $Q_{i}'$. The equations for
$Q_{i}$ are multiplied by $\phi_{i}'$ and the equations for $Q_{i}'$
are multiplied by $\phi_{i}$. Summing the two sets of equations then
gives eq. \ref{eq:29}.

The capacitance calculations we have discussed can also be applied
to finding the energy and forces on conductors and to calculating
transmission line impedances. The energy of a set of $N$ conductors
with charges $Q_{i}$ and potentials $\phi_{i}$ is \cite{jackson1975}\begin{equation}
W=\frac{1}{2}\sum_{i=1}^{N}Q_{i}\,\phi_{i}=\frac{1}{2}\sum_{i=1}^{N}\sum_{j=1}^{N}c_{ij}\,\phi_{i}\,\phi_{j}\label{eq:30}\end{equation}
To calculate the force on a conductor in a given direction, with all
conductors held at constant potential, we displace the conductor in
that direction and then calculate the new $c_{ij}'s$ and the new
energy. The force is then the change in energy divided by the lattice
constant.

The impedance of the transmission line formed by two conductors is
given by the equation\begin{equation}
Z=\frac{\eta}{C}\label{eq:31}\end{equation}
where $\eta$ is the characteristic impedance of the medium, given
by $\eta=\sqrt{\mu/\epsilon}$, and $C$ is the capacitance given
by eq. \ref{eq:21}. Note that $C$ is treated here as a pure number
i.e. it is not scaled by $\epsilon$.

Formulating electrostatics problems and their solution entirely in
discrete space has many advantages. This was first recognized in a
well known paper by Courant et al \cite{courant67} in which they
examined the solution of elliptic partial differential equations and
their corresponding difference equations. They were able to show that
the difference equation solution does converge to the solution of
the differential equation and that some questions, such as the existence
of solutions, are more easily answered by looking at the difference
equation. It appears that in many cases the theorems and methods developed
for continuous space problems can be translated over into discrete
space. It also seems to us that the possibility exists for discovering
new theorems or generalizations of existing theorems, such as Thompson-Lampard
\cite{thompson56}, by approaching problems from a discrete space
point of view. Opportunities for more research in this area certainly
exist.

\begin{acknowledgments}
The authors acknowledge the generous support of Exstrom Laboratories
and its president Istvan Hollos.

\bibliographystyle{apsrev}
\bibliography{/doc/articles/lgfint/gf,caplgf}

\end{acknowledgments}
\begin{figure}[p]
\includegraphics{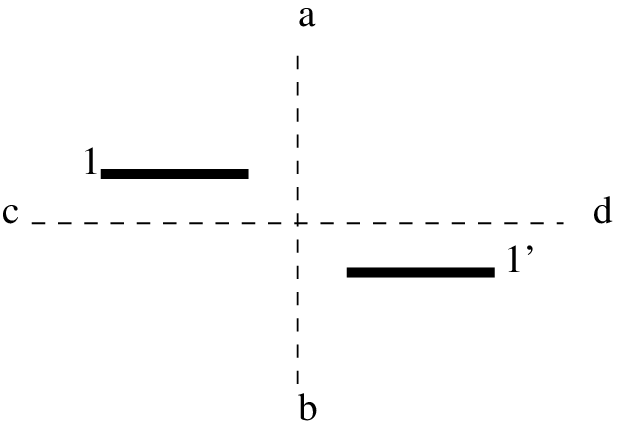}

\caption{\label{fig:1}Symmetry example.}
\end{figure}
\begin{figure}[p]
\includegraphics{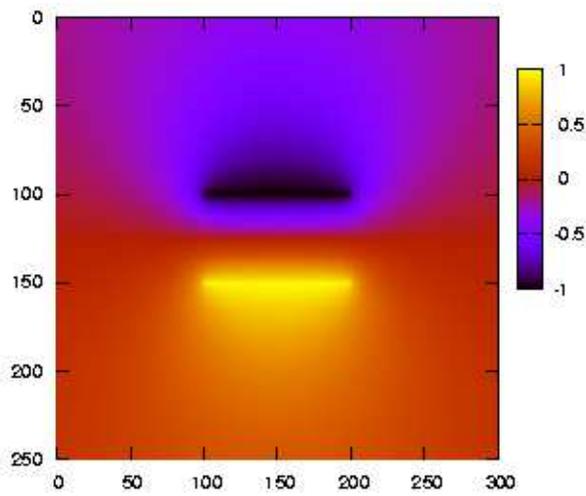}

\caption{\label{fig:2}Parallel stripline potential: 100 charges/plate, separation
is 1/2 plate width.}
\end{figure}
\begin{figure}[p]
\includegraphics[%
  scale=0.5]{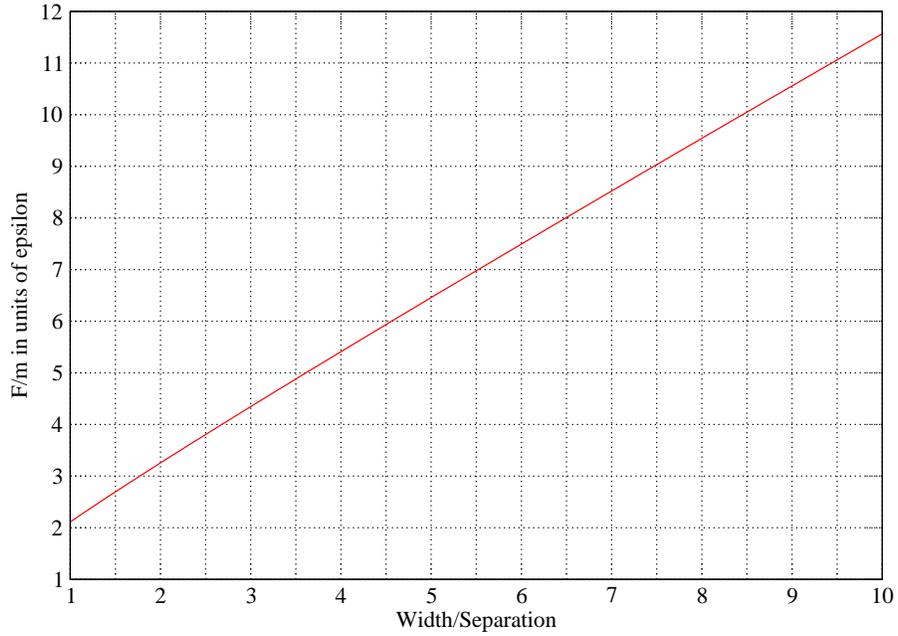}

\caption{\label{fig:3}Parallel stripline capacitance: F/m in units of epsilon
vs. width/separation. 1000 charges per plate.}
\end{figure}
\begin{figure}[p]
\includegraphics[%
  scale=0.5]{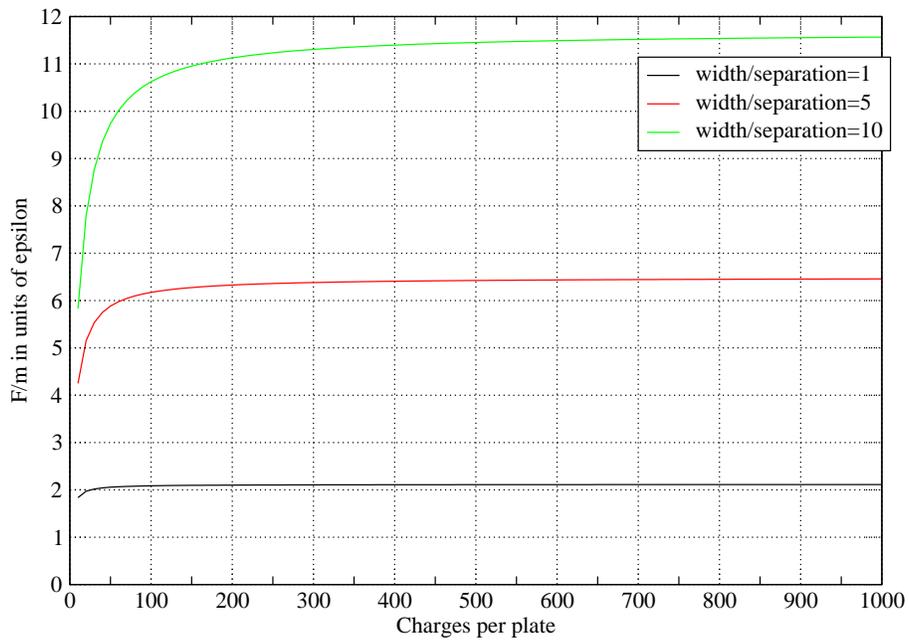}

\caption{\label{fig:4}Parallel plate convergence: F/m in units of epsilon
vs. charges per plate}
\end{figure}
\begin{figure}[p]
\includegraphics{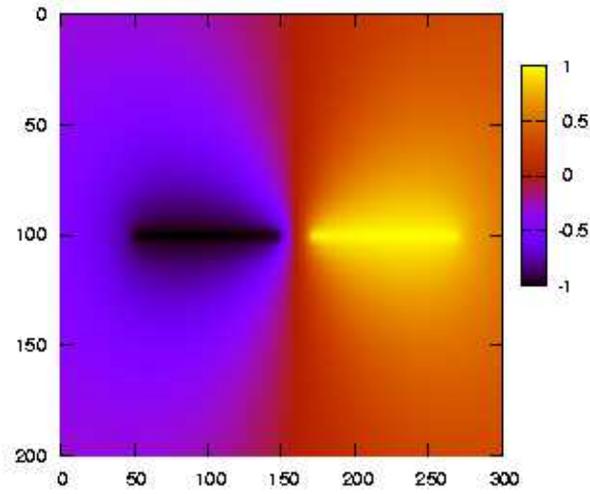}

\caption{\label{fig:5}Coplanar stripline potential: 100 charges/plate, separation
is 1/5 plate width.}
\end{figure}
\begin{figure}[p]
\includegraphics[%
  scale=0.5]{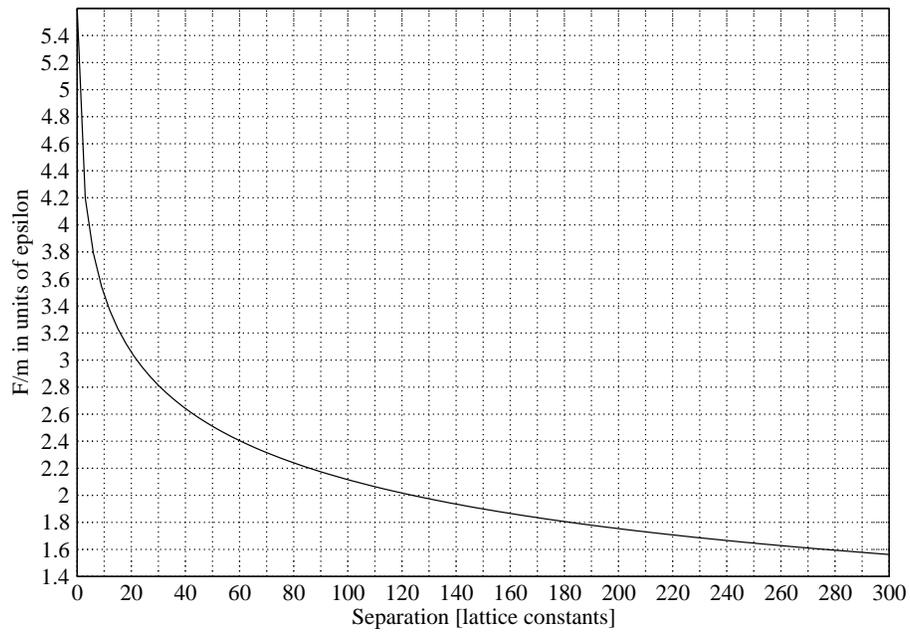}

\caption{\label{fig:6}Coplanar stripline capacitance: F/m in units of epsilon
vs. separation in units of lattice constants. 300 charges per plate.}
\end{figure}

\end{document}